\documentclass[aps,prb,twocolumn,groupedaddress,showpacs,floatfix]{revtex4}
\usepackage{epsfig}
\usepackage{epsf} 
\begin{document}
%

\title{Ab-initio calculation of phonon dispersion curves: accelerating
  q point convergence}

\author{Katalin Ga\'al-Nagy$^{1,2}$}
\email{katalin.gaal-nagy@physik.uni-r.de}
\author{Dieter Strauch$^{1}$}
\affiliation{$^{1}$Institut f{\"u}r theoretische Physik,
  Universit{\"a}t
  Regensburg, D-93040 Regensburg, Germany\\
  $^{2}$European Theoretical Spectroscopy Facility (ETSF),
  CNISM-CNR-INFM, and
  Dipartimento di Fisica dell'Universit\`a degli Studi
  di Milano, via Celoria 16, I-20133 Milano, Italy}

\date{\today}
%
\begin{abstract}
We present a scheme for the improved description of the long-range
interatomic force constants in a more accurate way than the procedure
which is commonly used within plane-wave based density-functional
perturbation-theory calculations. Our scheme is based on the
inclusion of a {\bf q} point grid which is denser in a restricted
area around the center of the Brillouin Zone than in the remaining
parts, even though the method is not limited to an area around
$\Gamma$. We have tested the validity of our procedure in the case of
high-pressure phases of bulk silicon considering the bct and sh
structure.
\end{abstract}
\maketitle
%
%
\section{Introduction}\label{intro}
The calculation of vibrational properties of solids, surfaces, and
nanocrystals plays an important role in the structural
characterization of matter. Besides the possibility to compare the
vibrational frequencies with experimental data, phonon modes can be
used also to determine the structural stability of a
system.\cite{Zha06, Li06, Xia05, Gaa06a} This requires a reliable
description of the vibrational properties of a system. Nowadays,
ab-initio methods based on the density-functional perturbation theory
(DFPT)\cite{Bar01} or frozen-phonon techniques\cite{Ihm81} are
commonly used for this purpose. With the latter the phonon frequencies
can be obtained by calculating the dynamical matrix for each {\bf q}
point along the high-symmetry directions of the Brillouin zone (BZ),
which can quite cumbersome. In the former case the dynamical matrices
are computed just on a (finite) grid of {\bf q} points in the
irreducible wedge of the Brillouin zone (IBZ) with subsequent Fourier
interpolation. Because of the computational effort of these
calculations, the grid is often taken as small as possible resulting
sometimes in an inaccurate description of the long-range force
constants yielding erroneous frequencies especially in the
low-frequency range near the $\Gamma$ point. Nevertheless, also other
frequencies in the Brillouin Zone can be affected.

Considering the investigation of the thermodynamic stability of the
system, an error in the description of the low-frequency area close to
$\Gamma$ plays just a minor role for the calculation of the free
energy. However, for the investigation of the Gr{\"u}neisen parameters or
the thermal expansion especially at low temperatures it is necessary
to describe particularly these frequencies correctly, since their
reciprocal value enter the formula.\cite{Ash76} Furthermore, if
someone is interested in the determination of soft phonon modes, which
point at the instability of structure, an exact description of the
corresponding frequencies is necessary. In this context one has to
mention that a wrong description of the long-range force constants may
yield imaginary frequencies for a stable structure, which are
interpreted as a result of a structural instability.

A possibility to overcome this problem has been given by Gonze and
Lee\cite{Gon97} by correcting the long-range dipole-dipole interaction
contribution to the force constants using a term which yields the
correct non analytic behavior in the limit of small {\bf q} similar to
the non analytic corrections yielding the LO-TO splitting. With this
procedure, the description of phonons close to $\Gamma$ is
significantly improved and the required number of {\bf q} points is
reduced. However, this scheme is restricted to semiconductors and can
not be employed in case of discrepancies at points close to the
boundary of the BZ, too, at least this possibility has not been proven
yet.

We have taken another approach to overcome the problem of the correct
description of the long-range force constants within DFPT, which is
related to the {\bf q}-point convergence by introducing a
mini-Brillouin Zone (mini-BZ). Inside of this mini-BZ the dynamical
matrices are computed for a denser mesh of {\bf q} points. The mini-BZ
can be chosen around the center of the BZ but also at its
boundary. These additional contributions are taken into account in the
determination of the force constants. The validity of this procedure
has been verified in the case of the high-pressure phases of bulk
silicon: we have chosen the body-centered tetragonal structure (bct)
corresponding to the $\beta$-tin and the simple hexagonal (sh)
structure corresponding to the sh phase. In both cases structures
beyond the range of structural stability of the phase have been
selected. Our method yields an improved description of the phonon
dispersion curves using less {\bf q} points than the standard
refinement of the grid. In detail, imaginary frequencies for the bct
structure near $\Gamma$ standard procedures have been traced back to
erroneous force constants whereas the soft phonon mode for the sh
structure have been verified.

This article is organized as follows: In Sect.~\ref{theory} we
describe the theoretical framework of our method. Next, a short review
of the technical details of our investigation is given
(Sect.~\ref{method}). In Sect.~\ref{results} we apply our procedure to
the bct (Sect.~\ref{resultsBCT}) and the sh (Sect.~\ref{resultsSH})
structure of bulk silicon, where the results are compared and
discussed afterwards in Sect.~\ref{resultsDiscussion}. Finally we
summarize and draw a conclusion (Sect.~\ref{conclusion}).
%
%
\section{Theory}\label{theory}
In general, the phonon frequencies $\omega ({\bf q})$ at a given {\bf q}
point in the BZ are obtained by diagonalizing the dynamical matrix
${\bf D}_{\alpha \alpha'}(\kappa \kappa', {\bf q})$, 
i.e., by solving the equation
\begin{eqnarray} \label{Eq_diag}
  \left\| \ 
  {\rm \bf D}_{\alpha \alpha'}{{\rm \bf q}, \kappa \ \kappa'} 
  - \omega^2({\rm \bf q})\ {\bf 1}_{3N_{\rm A}}\  \right\| = 0 \quad, 
\end{eqnarray}
where $\kappa$, $\kappa'$ label the sublattices (the basis atoms),
$\alpha,\alpha'$ are the cartesian coordinates, and ${\bf 1}_{3N_{\rm
A}}$ is the $3N_{\rm A} \times 3N_{\rm A}$ unitary matrix for $N_{\rm
A}$ atoms in the cell. We focus on the Fourier-interpolation scheme to
obtain the frequencies along the high-symmetry directions of the BZ.
To this end, DFPT calculations of the dynamical matrices are performed
for {\bf q} points on a finite, regular grid of {\bf q} points. These
dynamical matrices are connected with the force constants matrices
$\Phi_{\alpha \alpha'}{l \ l' \choose \kappa \ \kappa'}$
by a discrete Fourier transform (FT):
\\ \parbox{7.5cm}{\begin{eqnarray*}
  {\bf D}_{\alpha \alpha'}(\kappa \kappa', {\bf q}) &=& 
  \sum_{l l'}
  \Phi_{\alpha \alpha'}{l \ l' \choose \kappa \ \kappa'}
  \times \\
  & & \exp \left\{ -i {\bf q} \cdot
  \left[ 
    {\bf R} {l \choose \kappa} - {\bf R} {l' \choose \kappa'}
    \right] \right\} \quad ,
 \end{eqnarray*}}\hfill
\parbox{8mm}{ \begin{eqnarray} \end{eqnarray}} \\
where ${\bf R}(l,\kappa)$ is the coordinate of the $\kappa$th atom in
the $l$th cell. Using the common procedure within plane-wave codes one
calculates the force constants by a FT from the dynamical matrices for
the {\bf q} points from the discrete and finite grid and subsequently
obtains the dynamical matrices for any {\bf q} point by FT of the
force constant matrices. However, the drawback of this procedure is
that the range of the forces is connected with the {\bf q} points: the
choice a $l_1$$\times$$l_2$$\times$$l_3$ grid of {\bf q} points leads
to the inclusion of interatomic force constants between atoms within
$l_1$$\times$$l_2$$\times$$l_3$ cells. In other words, assuming a
finer {\bf q} point grid one can extend the range of the forces
included in the force constants. Thus, usually convergence tests have
to be performed comparing the phonon dispersion curves for
calculations based on various grids. However, the number of {\bf q}
points is $l_1\cdot l_2\cdot l_3$ in the BZ (which can be reduced by
symmetry), and therefore most investigations are restricted to a
smaller set of {\bf q} points with a less accurate description of the
long-range force constants.

The long-range force constants affect particularly the low-frequency
phonons close to the $\Gamma$ point which might be wrongly described
within this procedure. This problem can be overcome for semiconductors
by using the method described in Ref.~[\onlinecite{Gon97}]. Another
possibility is to use more {\bf q} points in the area close to
$\Gamma$ which we have chosen here.

Instead of taking increasingly grids we have taken a denser grid just
in a small area around $\Gamma$ (mini-BZ), and we assume a less dense
grid outside this mini-BZ. Since the FT is based on a regular grid, we
interpolate the missing dynamical matrices outside the mini-BZ by a FT
from the force constants based on the coarser grid. In the following
detailed description we neglect the indices $\kappa$ and $\alpha$ for
simplicity, and we investigate the case $l_1=l_2=l_3$ only. The case
$l_1\not=l_2\not=l_3$ follows analogously.

Assuming an $n$$\times$$n$$\times$$n$ grid, the force constants
$\Phi^{nnn}$ are obtained by FT from the corresponding dynamical
matrices ${\bf D}^{nnn}$ calculated within DFPT:
\begin{eqnarray}\label{Eq_ForceTrue}
  {\bf D}^{nnn}({\bf q}) \stackrel{\rm FT} {\longrightarrow}
  \Phi^{nnn} \quad .
\end{eqnarray}
From these force constants $\Phi^{nnn}$ the dynamical matrices for any
{\bf q} point in the BZ can be calculated by a back FT, therefore also
for the {\bf q} points on a finer grid, e.g., a
$l$$\times$$l$$\times$$l$ grid with $l>n$. Thus, one gets
\begin{eqnarray}
  \Phi^{nnn} \stackrel{\rm FT} {\longrightarrow} {\bf D}_{\rm
    int}^{lll}({\bf q}) \quad ,
\end{eqnarray}
and also from these dynamical matrices ${\bf D}_{\rm int}^{lll}({\bf
q})$ one can get again force constants $\tilde{\Phi}_{\rm int}^{lll}$
by
\begin{eqnarray}\label{Eq_ForceInt}
  {\bf D}^{lll}_{\rm int}({\bf q}) \stackrel{\rm FT} {\longrightarrow}
  \tilde{\Phi}^{lll}_{\rm int} \quad ,
\end{eqnarray}
which are in this case identical to $\Phi^{nnn}$. In a next step we
have performed calculations of the dynamical matrices ${\bf
D}^{lll}({\bf q}\in$mini-BZ$)$ for {\bf q} points of the
$l$$\times$$l$$\times$$l$ grid inside the mini-BZ using the DFPT
procedure. Taking these dynamical matrices and the interpolated ones
outside the mini-BZ, the improved force constants $\Phi^{lll}_{\rm
int}$ can be calculated by
\begin{eqnarray}
  {\bf D}^{lll}({\bf q}\in\mbox{\rm mini-BZ}) \vee 
  {\bf D}_{\rm int}^{lll}({\bf q}\not\in\mbox{\rm mini-BZ})
  \stackrel{\rm FT} {\longrightarrow} \Phi^{lll}_{\rm int} \quad .
\end{eqnarray}
From these force constants $\Phi^{lll}_{\rm int}$ the dynamical
matrices for any {\bf q}, in particular along the high-symmetry
directions of the BZ, are calculated in the standard way. In the limit
of very fine grids one should achieve the same results as within the
procedure of Gonze and Lee.\cite{Gon97} However, this method is not
restricted to a mini-BZ around the $\Gamma$ point since the mini-BZ
can be chosen arbitrarily. This has the advantage that also ranges of
the dispersion curves far away from the $\Gamma$ point can be
improved. Besides, the shape of the phonon density of states for
particular spectral features can be inspected in detail.

In order to apply this scheme we have modified a postprocessing
routine of the QUANTUM ESPRESSO package\cite{pwscf} in order to write
out not only the frequencies but also the complete dynamical matrices
after the FT.  We have checked the numerical stability of the method
by comparing the phonon dispersion curves based on the force constants
$\Phi^{888}$ from Eq.~(\ref{Eq_ForceTrue}) and $\tilde{\Phi}_{\rm
int}^{888}$ from Eq.~(\ref{Eq_ForceInt}), and we have found
differences in the frequencies of less than $0.25~{\rm cm}^{-1}$.

In the following we apply the procedure for testing purpose to the
two silicon structures mentioned above and describe how to choose the
mini-BZ and the required $l$$\times$$l$$\times$$l$ grid.
%
%
\section{Method}\label{method}

All calculations have been carried out with the QUANTUM ESPRESSO
package.\cite{pwscf} It is based on a plane-wave pseudopotential
approach to the density-functional theory (DFT).\cite{Hoh64, Koh65}
For silicon we have employed a norm-conserving pseudopotential
generated following the scheme suggested by v.\ Barth and
Car.\cite{CarUn,Cor93} The exchange-correlation energy is described
within the local-density approximation (LDA).\cite{Per81, Cep80} We
have used a kinetic-energy cutoff of 40~Ry and a
20$\times$20$\times$20 Monkhorst-Pack mesh\cite{Mon76} together with a
Methfessel-Paxton smearing\cite{Met89} using a width of 0.03~Ry to
describe the electronic (metallic) ground state of the systems. The
phonon frequencies have been calculated using the DFPT
scheme\cite{Bar87, Gia91, Gir95} as implemented in the QUANTUM
ESPRESSO package followed by a discrete Fourier Transform as described
in Sect.~\ref{theory}.

Both, the sh and bct structures have been investigated using a common
body-centered orthorhombic cell (bco, lattice constants $a \not= b
\not= c$) with two atoms at $(0,0,0)$ and at $(0,0.5b,\Delta c)$ in
the unit cell. The symmetry of bct requires $a=b$ and $\Delta=0.25$
whereas the symmetry of sh yields $b=\sqrt{3}c$ and $\Delta=0.5$. In
fact, for sh we use a biatomic supercell although the structure of the
sh phase can be described with just one atom in the sh unit
cell. However, using the bco cell we have access to soft modes
corresponding to the doubling of the unit cell. For details of the
choice of the cell see Ref.~[\onlinecite{Gaa06a}]. In this work, we
have relaxed the ground-state geometry of the structure for a volume
fixed to 184~$a_{\rm B}^{3}$ for both sh and bct. The equilibrium
lattice constants are $c/a=0.5489$ for bct and $c/a=0.5338$ and
$b/a=0.9230$ for sh. The error with respect to the ideal $b/c$ ratio
of sh is negligible.

%
\section{Results}\label{results}
For the application of our method we have chosen the bct structure of
silicon at $V=184~a_B^3$ which is a volume beyond the stability range
of the corresponding $\beta$-tin phase. For this structure we have
found a phonon instability along the $\Gamma$-X direction of the bct
BZ\cite{Gaa99,Gaa01} which is equivalent to the $\Gamma$-T direction
of the bco BZ (see Ref.~[\onlinecite{Gaa06a}]). This phonon
instability turned out not to be physical. The phonons in the
mentioned articles had been calculated using a 4$\times$4$\times$4
Monkhorst-Pack mesh, which was slightly insufficient to describe the
frequencies in this region of the BZ properly, since calculations
within DFPT of dynamical matrices at {\bf q} points near $\Gamma$ have
yielded only real frequencies. Because the phase space of the
numerically soft modes was negligibly small, the imaginary frequencies
did not affect the results for the free energy.  However, we use this
case to check the validity of the method described in
Sect.~\ref{theory} by the use of the mini-BZ based on a
4$\times$4$\times$4 grid outside and a 8$\times$8$\times$8 one inside
the mini-BZ. A comparison of the phonon dispersion curve using the
4$\times$4$\times$4 grid plus the mini-BZ with to the one based on
8$\times$8$\times$8 mesh gives an estimate of the errors using our
scheme. The bct structure is also taken to exemplify the choice of the
extent of the mini-BZ.  Then we apply our scheme to go beyond the
8$\times$8$\times$8 mesh for the final results.

As a second example we have chosen the sh structure of silicon at
$V=184~a_B^3$ which is also a volume beyond the stability range of the
corresponding phase, here the sh phase. Choosing a biatomic supercell,
a soft phonon mode has been found at the $\Gamma$ and the S point of
the bco BZ.\cite{Gaa06a} Both points are equivalent to the $\Gamma$
point of the monatomic sh unit cell. The finding of a soft phonon mode
at the $\Gamma$ point is in accordance with other reported
results.\cite{Cha84,Nee84,Cha85} In fact, the softening at the
BZ-boundary point S refer to a doubling of the sh unit cell. This is
in agreement with the distortion of the biatomic supercell which
contains two monatomic sh cells. Thus, this soft phonon mode is of
physical origin.

Both case studies, the one with an unphysical but numerically
imaginary frequencies and the one with the physically correct phonon
instability are described in the following, and they are finally
compared and discussed.
%
%
\subsection{Application to the bct structure of silicon}\label{resultsBCT}
%
%
\subsubsection{Standard procedure: Increasing mesh size}

\begin{figure}[t]
  \begin{minipage}{8.6cm} 
    \epsfig{figure=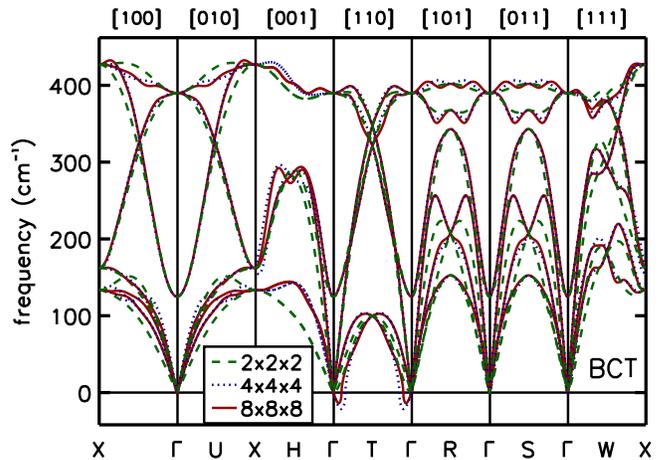,width=8.6cm,angle=0}
  \end{minipage}
  \caption{(Color online) Phonon dispersion curves for the bct
    structure at $V=184~{\rm a_B^3}$. 
    Dispersion curves obtained by FT using a 2$\times$2$\times$2 mesh
    (dashed lines), 4$\times$4$\times$4 mesh (dotted lines), and a
    8$\times$8$\times$8 mesh (solid lines).
  }\label{Pic_BctDisp}
\end{figure}
For the bct structure at $V=184~a_B^3$ corresponding to a pressure of
133~kbar the frequencies along the high-symmetry directions of the bco
BZ have been calculated using a 2$\times$2$\times$2, a
4$\times$4$\times$4, and a 8$\times$8$\times$8 grid. The results are
shown in Fig.~\ref{Pic_BctDisp}. Since the bct structure has a higher
symmetry than the bco structure, the $\Gamma$-X and the $\Gamma$-R
directions (bco) are equivalent to the $\Gamma$-U-X and the $\Gamma$-S
directions (bct), respectively.  The equivalent directions are shown
mainly for completeness. The $\Gamma$-T direction with
T$(\frac{1}{2},\frac{1}{2},0)$ (coordinates in units of reciprocal
lattice vectors; in the following we will assume the points in the BZ
always in units of reciprocal lattice vectors without mentioning it
explicitly) is of particular interest because a soft phonon mode
appears close to $\Gamma$ using the 4$\times$4$\times$4 mesh as
visible in Fig.~\ref{Pic_BctDisp}. Note, that this softening does not
appear for the 2$\times$2$\times$2 grid, which is obviously
insufficient to describe the phonon dispersion correctly. Increasing
the mesh size, the softening remains for a 8$\times$8$\times$8 grid,
however, to a minor extent. From the dispersion curves it is difficult
to decide whether convergence with respect to {\bf q} has been
achieved for the 4$\times$4$\times$4 or not, since the the frequencies
at the mesh points of the 8$\times$8$\times$8 grid are on top of the
Fourier-interpolated 4$\times$4$\times$4 dispersion curves. Only for
the low-frequency mode along the $\Gamma$-T direction some of the
interpolated frequencies are imaginary but the 8$\times$8$\times$8
points yield just real values for the frequencies. Besides, the shape
of the 8$\times$8$\times$8 Fourier-interpolated phonon curves shows
just minor deviations from the 4$\times$4$\times$4 ones. Inspecting
the frequencies at mesh points of the 16$\times$16$\times$16 grid, the
calculated frequencies are nearly indistinguishable from the
interpolated dispersion curves except along the critical $\Gamma$-T
direction where all calculated points yield real frequencies, while a
part of the 8$\times$8$\times$8 Fourier-interpolated phonon dispersion
are imaginary.

\begin{figure}[t]
  \begin{minipage}{8.6cm} 
    \epsfig{figure=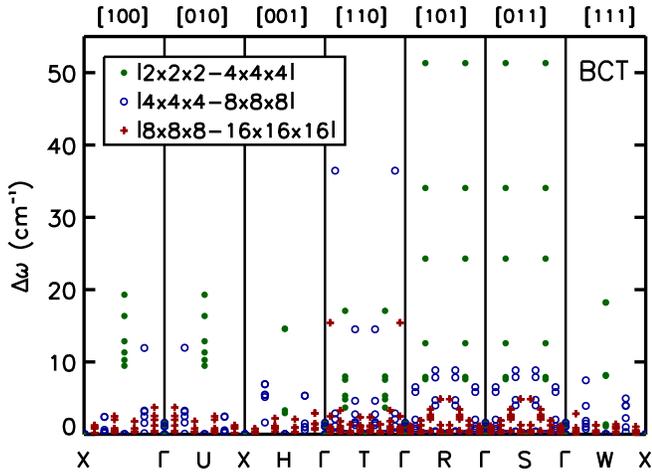,width=8.6cm,angle=0}
  \end{minipage}
  \caption{(Color online) Differences $\Delta \omega$ between the
    calculated and Fourier interpolated points along the high-symmetry
    directions for various meshes: 
    $\Delta \omega = |\omega_{222}^{\rm int}-\omega_{444}^{\rm DFPT}|$ 
    are drawn with solid symbols, 
    $\Delta \omega = |\omega_{444}^{\rm int}-\omega_{888}^{\rm DFPT}|$ 
    with open symbols,  and
    $\Delta \omega = |\omega_{888}^{\rm int}-\omega_{16\,16\,16}^{\rm DFPT}|$ 
    are with crosses (see text). 
  }\label{Pic_BctDiff}
\end{figure}
However, such a detailed study is not always possible for every
system. In our case, the 2$\times$2$\times$2 mesh required the
calculation of dynamical matrices at 4 {\bf q} points in the IBZ, the
4$\times$4$\times$4 mesh at 13 {\bf q} points, and the
8$\times$8$\times$8 mesh at 59 {\bf q} points. Since the dispersion
curves do not change significantly assuming a grid denser than the
4$\times$4$\times$4 except along the $\Gamma$-T direction and there
only in the region close to $\Gamma$ it is not necessary to calculate
all the dynamical matrices on a 8$\times$8$\times$8 or a
16$\times$16$\times$16 grid. Ultimately, the unphysical
instability should be lifted. This can be achieved by applying our
method described in Sect.~\ref{theory} using a mini-BZ around
$\Gamma$. The extent of the mini-BZ can be determined by inspecting the
differences between the frequencies derived from the DFPT dynamical
matrices $\omega^{\rm DFPT}$ and the Fourier-interpolated ones
$\omega^{\rm int}$. For this purpose we have drawn in
Fig.~\ref{Pic_BctDiff} the differences
\begin{eqnarray}\label{Eq_diff}
  \Delta \omega = |\omega_{nnn}^{\rm int}-\omega_{lll}^{\rm DFPT}|
  \quad ,
\end{eqnarray}
using an $l$$\times$$l$$\times$$l$ and an $n$$\times$$n$$\times$$n$ grid
with $l>n$ for the frequencies presented in Fig.~\ref{Pic_BctDisp}. As
mentioned above, the differences $\Delta\omega$ for the
2$\times$2$\times$2 and the 4$\times$4$\times$4 mesh are quite
large. All differences decrease significantly using a finer grid,
except for the points near $\Gamma$ along the $\Gamma$-T
direction. Thus, the application of our method for a mini-BZ around
$\Gamma$ promises an improvement of the results especially in the
range of $\Gamma$-T.

%
%
\subsubsection{Approval of the present Mini-BZ method}

For a first test we want to apply our method using a
4$\times$4$\times$4 grid outside the mini-BZ and an
8$\times$8$\times$8 inside. The scope of this test is to reproduce the
8$\times$8$\times$8 curves (inclusive the numerically instable mode)
using a 4$\times$4$\times$4 mesh together with a Mini-BZ, since a
phonon dispersion curve using a full 8$\times$8$\times$8 grid exists
as a reference.

Inspecting Fig.~\ref{Pic_BctDiff}, the largest deviation of the
8$\times$8$\times$8 mesh from the 4$\times$4$\times$4 mesh (open
symbols in Fig.~\ref{Pic_BctDiff}) of $\Delta \omega \approx 36~{\rm
cm}^{-1}$ is found for the point $(\frac{1}{8},\frac{1}{8},0)$ along
$\Gamma$-T, but also for the point $(\frac{3}{8},\frac{3}{8},0)$ the
differences between the meshes are in the order of $\Delta \omega
\approx 15~{\rm cm}^{-1}$. Besides, along X-$\Gamma$-U-X the
differences are also remarkable for {\bf q} points with components
$q_i\leq\frac{1}{4}\ (i=x,y,z)$. Thus, the selection of {\bf q} points
up to $q_i\leq\frac{3}{8}$ would be a promising choice for the
mini-BZ.

\begin{figure}[t]
  \begin{minipage}{8.6cm} 
    \epsfig{figure=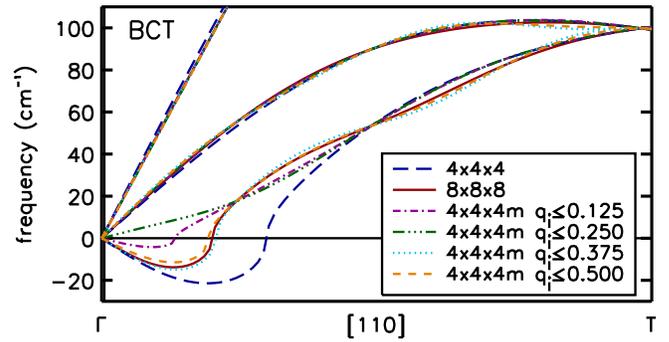,width=8.6cm,angle=0}
  \end{minipage}
  \caption{(Color online) Phonon dispersion curves along $\Gamma$-T
    using a 4$\times$4$\times$4 and a 8$\times$8$\times$8 mesh
    together with interpolated curves (4$\times$4$\times$4m) based on
    various mini BZs as denoted in the inset. Imaginary frequencies
    are drawn along the negative frequency axis. The cutoff for the
    {\bf q} points is in units of reciprocal lattice vectors for $q_i$
    with $i=x,y,z$ (see text).
  }\label{Pic_BctMiniBZ}
\end{figure}
\begin{figure}[t]
  \begin{minipage}{8.6cm} 
    \epsfig{figure=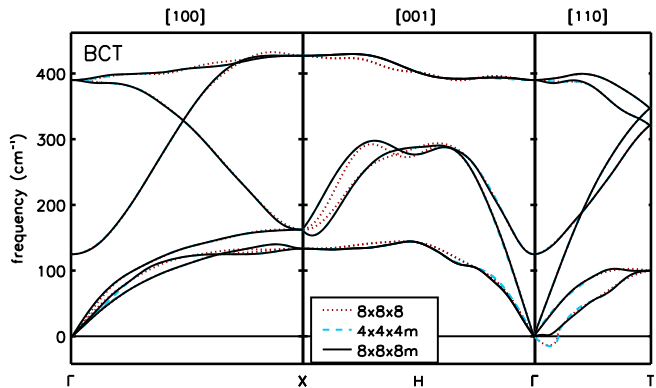,width=8.6cm,angle=0}
  \end{minipage}
  \caption{(color online) Comparison of the dispersion obtained from a
    8$\times$8$\times$8 grid (dotted line), using the $q_i\leq0.375$
    mini BZ in addition to the 4$\times$4$\times$4 grid
    (4$\times$4$\times$4m, dashed line), and using
    16$\times$16$\times$16 points in the $q_i\leq0.25$ mini BZ based
    on force constants from the 4$\times$4$\times$4m grid
    (8$\times$8$\times$8m, solid line), see text. Imaginary
    frequencies are drawn along the negative frequency axis.
  }\label{Pic_BctAppl}
\end{figure}

In the following we have used various mini-BZs up to
$q_i\leq\frac{1}{2}\ (i=x,y,z)$ for the interpolation (denoted as
4$\times$4$\times$4m). The results for the low-frequency range of the
phonon dispersion curve along the $\Gamma$-T direction are shown in
Fig.~\ref{Pic_BctMiniBZ} in comparison with the results based on
4$\times$4$\times$4 and 8$\times$8$\times$8 grids.  Note: the choice
of $q_i\leq1$ would make the 4$\times$4$\times$4m grid identical to
the 8$\times$8$\times$8 grid.  The curves for the mini-BZ with
$q_i\leq\frac{3}{8}$ and $q_i\leq\frac{1}{2}$ are both very close to
the curve using the full 8$\times$8$\times$8 mesh. Therefore, the
choice of $q_i\leq\frac{3}{8}$ for the mini-BZ which has been already
assumed from Fig.~\ref{Pic_BctDiff}, is confirmed. Also the
high-frequency region of the dispersion using an 8$\times$8$\times$8
grid is reproduced very well with this 4$\times$4$\times$4m grid (see
Fig.~\ref{Pic_BctAppl}). Only minor deviations appear at regions more
distant from $\Gamma$ resulting from the unresolved $\Delta\omega =
6.93~{\rm cm}^{-1}$ along X-H-$\Gamma$. However, the general
improvement of the accuracy of the phonon-dispersion curve is
remarkable.  Note, that with this mini-BZ only 15~{\bf q} points of
the 8$\times$8$\times$8 have been necessary in addition to the 13~{\bf
q} points of the 4$\times$4$\times$4 mesh, which are much less than
the 59~{\bf q} points of the full 8$\times$8$\times$8 grid.

%
%
\subsubsection{Application of the Mini-BZ to denser grids}

After verifying the validity of our method we want to go beyond the
8$\times$8$\times$8 mesh because of the remaining unphysical softening
along $\Gamma$-T, whereas the calculated frequencies at
16$\times$16$\times$16 mesh points along the high-symmetry directions
show only real values. Inspecting the differences $\Delta\omega$ in
Fig.~\ref{Pic_BctDiff} again there is just a major difference of
$\approx 15 {\rm cm}^{-1}$ along $\Gamma$-T for {\bf
q}$=(\frac{1}{16},\frac{1}{16},0)$, whereas the other differences are
tiny. Since along the $\Gamma$-T direction at {\bf
q}$=(\frac{3}{16},\frac{3}{16},0)$ there is a crucial difference of
3.3~${\rm cm}^{-1}$, which might be important for resolving the
numerical soft mode, we have tested Mini-BZs with {\bf q} points up to
$q_i\leq\frac{1}{4}$ yielding additional 29 {\bf q} points.  The
results are denoted as 8$\times$8$\times$8m. With this mini-BZ the
phonon frequencies are described accurately as shown in
Fig.~\ref{Pic_BctAppl} and the softening along $\Gamma$-T has been
lifted.  The curves for 8$\times$8$\times$8m and 4$\times$4$\times$4m
match nearly exactly indicating that convergence has been achieved.
Note, that in this case the mini-BZ using 16$\times$16$\times$16
points is applied on top of the mini-BZ using 8$\times$8$\times$8
points in addition to the 4$\times$4$\times$4 mesh. In fact, we
achieve convergence for 8$\times$8$\times$8m since the remaining
differences $\Delta\omega$ are less than $1.25~{\rm cm}^{-1}$. It has
to be mentioned that further improvement could be achieved using a
mini-BZ close to the points T, R, and S, which is also possible within
our scheme.

In summary, we have been able to obtain converged phonon dispersion
curves using in total 57 {\bf q} points in the IBZ which is around
one sixth of the 349 {\bf q} points which are required for a
complete 16$\times$16$\times$16 mesh. In this way, the convergence is
accelerated significantly.

%
%
\subsection{Application to the sh structure of silicon}\label{resultsSH}
Similarly to bct we have investigated the sh structure of bulk
silicon at a volume of $V=184~{\rm a_B^3}$ which accords here to a
pressure of 107~kbar, again beyond the range of stability of the
corresponding sh phase.

%
%
\subsubsection{Standard procedure: Increasing mesh size}

\begin{figure}[h]
  \begin{minipage}{8.6cm} 
    \epsfig{figure=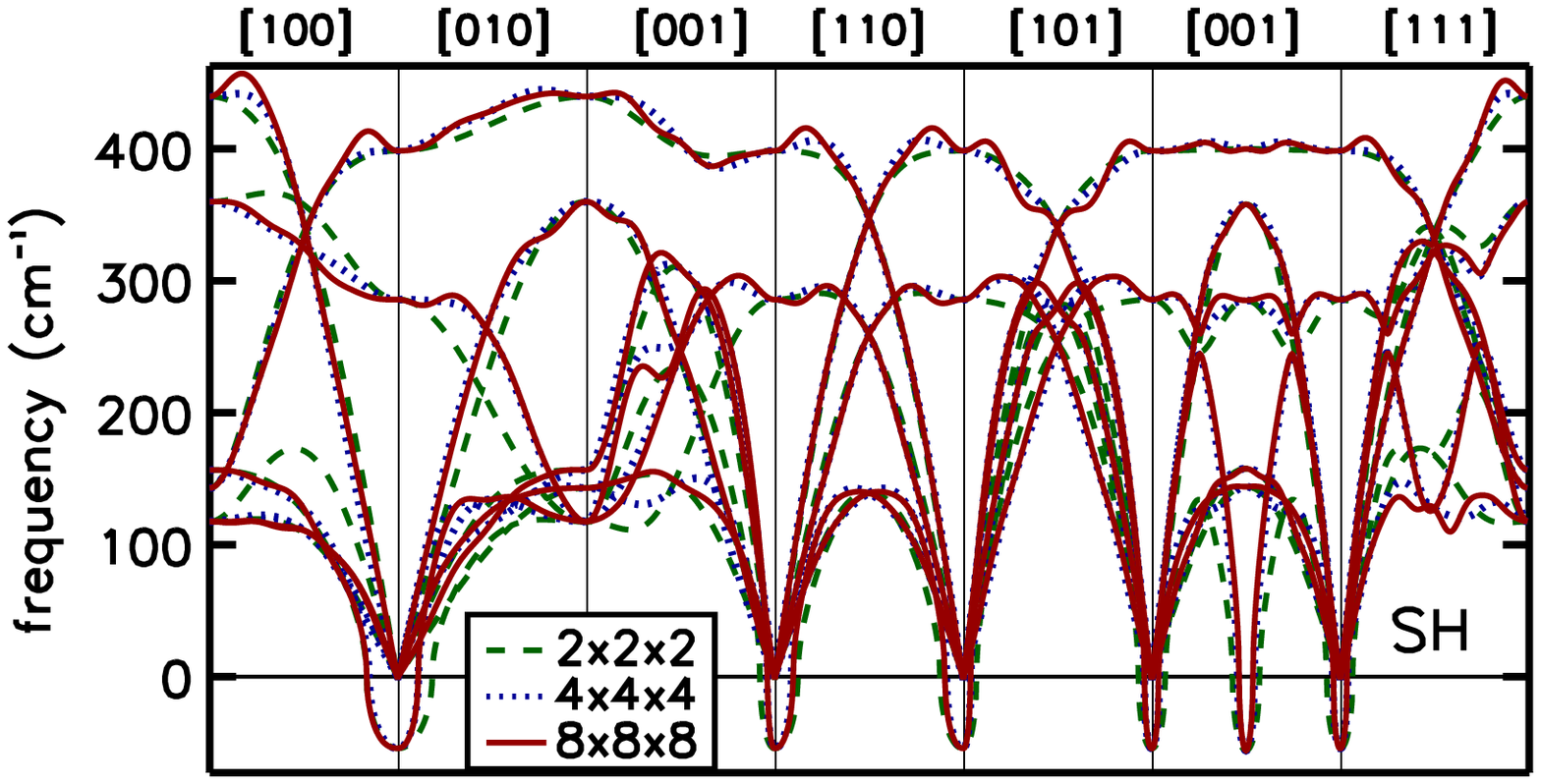,width=8.6cm,angle=0}
    \epsfig{figure=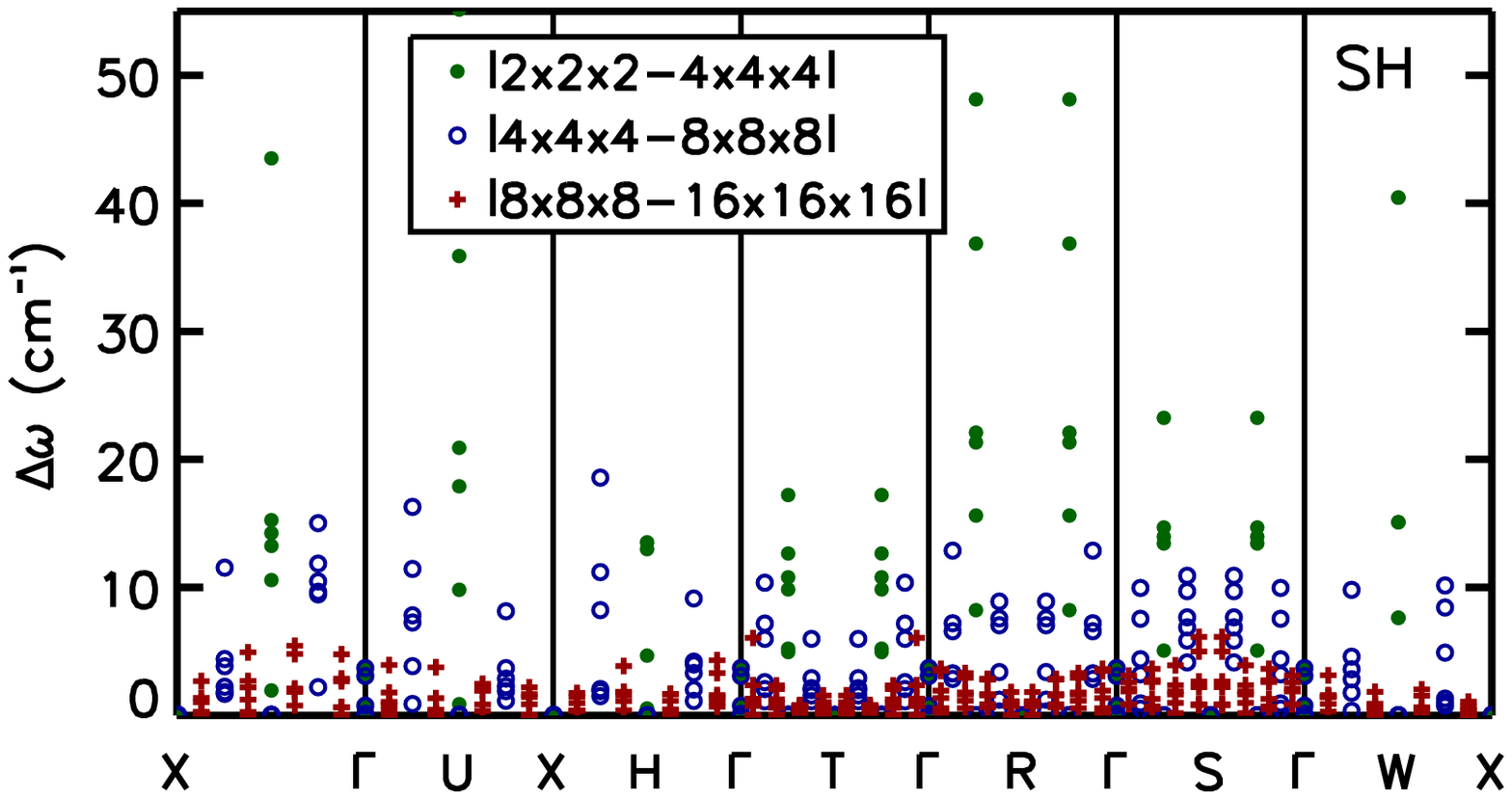,width=8.6cm,angle=0}
  \end{minipage}
  \caption{(color online) Phonon dispersion curves for the sh
    structure at $V=184~{\rm a_B^3}$. Upper panel: curves obtained
    from a 2$\times$2$\times$2 mesh (dashed lines), a
    4$\times$4$\times$4 mesh (dotted lines), and a 8$\times$8$\times$8
    mesh (solid lines). Imaginary frequencies are drawn along the
    negative frequency axis. Lower panel: differences $\Delta\omega$
    for sh as in Fig.~\ref{Pic_BctDiff}.
  }\label{Pic_SHDisp}
\end{figure}

First, we compare the phonon-dispersion curves using the
2$\times$2$\times$2, the 4$\times$4$\times$4, and the
8$\times$8$\times$8 grid and the differences $\Delta\omega$ (see
Eq.(\ref{Eq_diff})) in Fig.~\ref{Pic_SHDisp}. Because of the lower
symmetry of the sh structure, more {\bf q} points for each considered
grid had to be calculated within DFPT: 5 points for
2$\times$2$\times$2, 18 for 4$\times$4$\times$4, 95 for
8$\times$8$\times$8 and 621 for 16$\times$16$\times$16. Inspecting
Fig.~\ref{Pic_SHDisp}, there are remarkable differences visible in the
results based on a 2$\times$2$\times$2 and a 4$\times$4$\times$4 mesh
and between the ones based on a 4$\times$4$\times$4 and a
8$\times$8$\times$8 mesh. This is reproduced in the graph of
$\Delta\omega$. However, the variations along the $\Gamma$-T direction
are rather small and the the imaginary frequencies nearly do not
change the extension. In addition to the $\Gamma$ point there are
significant differences around the X and the S points. Since the
$\Gamma$ point and the S point are in this case equivalent, an
improvement at $\Gamma$ will yield an improvement at S. Now we focus
on the area around the $\Gamma$ point.

%
%
\subsubsection{Second test of the present method}

\begin{figure}[t]
  \begin{minipage}{8.6cm} 
    \epsfig{figure=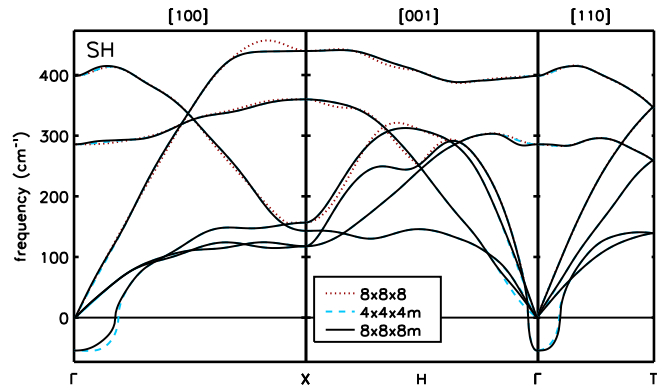,width=8.6cm,angle=0}
  \end{minipage}
  \caption{(color online) Comparison of the dispersion obtained from a
    8$\times$8$\times$8 grid (dotted line), using the $q_i\leq0.5$
    mini BZ for 8$\times$8$\times$8 points in addition to the
    4$\times$4$\times$4 grid (4$\times$4$\times$4m, dashed line), and
    using 16$\times$16$\times$16 points in the $q_i\leq0.125$ mini BZ
    in addition to 4$\times$4$\times$4m (see text). Imaginary
    frequencies are drawn along the negative frequency axis.
  }\label{Pic_ShAppl}
\end{figure}

Looking at the differences $\Delta\omega$ between the
4$\times$4$\times$4 and the 8$\times$8$\times$8 mesh, the choice of
$q_i\leq\frac{3}{8}$ as for bct is not reasonable for sh since the the
deviation of $\Delta \omega \leq 18.57~{\rm cm}^{-1}$ along the
X-H-$\Gamma$ direction at {\bf q}$=(0,0,\frac{7}{8})$ cannot be
reduced with a mini-BZ around $\Gamma$. Nevertheless, there are
differences of $\Delta\omega$ in the order of 15~${\rm cm}^{-1}$ close
to $\Gamma$ at {\bf q} with $q_i\leq\frac{1}{4}$ which can be
resolved. However, there are also significant deviations of $\approx
12~{\rm cm}^{-1}$ close to the R and S points. Thus we have chosen
$q_i\leq\frac{1}{2}$. With this mini-BZ in addition to the 18 {\bf q}
points of the 4$\times$4$\times$4 in further 40 dynamical matrices are
necessary. A comparison between the phonon dispersion curves based on
the full 8$\times$8$\times$8 mesh and the one using the mini-BZ
(4$\times$4$\times$4m) is presented in Fig.~\ref{Pic_ShAppl}. The
agreement is acceptable, only the overbending close to the X point is
not described correctly due to the choice of the mini-BZ around
$\Gamma$. Using an additional mini-BZ close to X would solve this
problem. However, the scope here is a check of the method for a
mini-BZ around $\Gamma$ analogously to the bct case
(Sect.~\ref{resultsBCT}), and this region is described excellently.

%
%
\subsubsection{Application of the Mini-BZ to go beyond the
8$\times$8$\times$8 grid}

Next, we want to go beyond the 8$\times$8$\times$8 mesh, again
focussing on the region around the center of the BZ. The differences
$\Delta \omega$ between the 8$\times$8$\times$8 and the
16$\times$16$\times$16 mesh show significant deviations for
$q_i\leq\frac{1}{8}$, with a maximum of $\Delta \omega \leq 6.13~{\rm
cm}^{-1}$ at the S point, but also for the $\Gamma$-X direction along
$[100]$. In order to reduce these discrepancies and for describing
especially the range of the phonon softening correctly, we have chosen
$q_i\leq\frac{1}{8}$ for the mini-BZ by including additional 10
dynamical matrices for the calculation of the force constants. In this
way, the differences $6.05~{\rm cm}^{-1}$ are eliminated along
$\Gamma$-T. The resulting phonon-dispersion curve is presented in
Fig.~\ref{Pic_ShAppl}. One can notice the improvement of the results
based on the 4$\times$4$\times$4m and the 8$\times$8$\times$8m
grid.

Considering the numerical effort, we have used 18 dynamical matrices
of the 4$\times$4$\times$4 mesh, additional 40 of the
8$\times$8$\times$8 one, and furthermore 10 of the
16$\times$16$\times$16 grid, in total 68, which is much less than the
621 dynamical matrices required for a complete 16$\times$16$\times$16
grid. Also in this case the numerical effort has been reduced
drastically.
%
%
\subsection{Discussion}\label{resultsDiscussion}
For both systems, the bct and the sh structure of bulk silicon, we
have been able to apply successfully our scheme and we have obtained
converged phonon-dispersion curves using less {\bf q} points than the
corresponding full mesh.  Comparing the results of the bct and the sh
structure, one can see that the convergence of the sh structure with
respect to the {\bf q} points is slower than the one of the bct
structure which results in a larger choice of the mini-BZ. In
particular, for bct the critical area with large variations with
respect to the choice of the {\bf q} mesh is around the center of the
BZ whereas for sh it is around X. Since the ground state of both
structures had been calculated using the same convergence parameters
and the same unit cell at the same volume, the different speed of the
{\bf q} convergence is not due to different convergence parameters for
the ground state. Therefore, one can not estimate the size of the {\bf
q} point grid which is required for convergence from one structure to
another one even using the same cell. Thus, the {\bf q} point
convergence has to be tested for any structure separately. However, it
was possible to confirm the soft phonon mode at $\Gamma$ for the sh
structure, whereas the one of the bct structure has disappeared
including enough {\bf q} points around the center of the BZ. Hence,
the latter phonon instability is due to an inaccurate description of
the long-range force constants and thus it has only numerical
origin. It would be possible to reduce the remaining discrepancies for
sh around the X point using our method by applying an additional
mini-BZ. However, this is beyond the scope of this article.

%
\section{Conclusions}\label{conclusion}
We have presented a scheme within standard DFPT calculations for the
accurate calculation of phonon dispersion curves by improving the
interatomic force constants which can be applied for semiconducting
systems as well as for metallic ones. Especially the long-range
contribution to the force constants can be described
successfully. This scheme is based on the inclusion of a denser {\bf
q}-point mesh in some part of the BZ (mini-BZ) and a wider one
outside. The method has been applied successfully to the bct and the
sh structure of bulk silicon, where the origin of a phonon instabilities
has been discussed. In detail, for the bct structure the soft phonon
mode has been traced back to an inaccurate description of the
(long-range) force constants and the imaginary frequencies become real
applying our procedure till convergence. In the case of the sh
structure the soft phonon mode has been confirmed. For both cases, the
number of required dynamical matrices has been reduced
drastically. Whereas here the mini-BZ has been chosen around the
center of the BZ, our scheme allows also a different choice. In this
way, also features at the boundary of the BZ can be described more
accurately. The use of our method can improve results not only for
phonon dispersion curves but also for related quantities like
Gr{\"u}neisen parameters, thermal expansion, and the free energy.
%
\section*{Acknowledgment}
Support by the Heinrich B\"oll Stiftung, Germany, is gratefully
acknowledged. Computer facilities at CINECA granted by INFM (Project
no. 643/2006) are gratefully acknowledged. This work was funded in
part by the EU's 6th Framework Programme through the NANOQUANTA
Network of Excellence (NMP-4-CT-2004-500198).
%
%
\bibliography{Theory,HighPressure,MyAndOthers}
%
%

\end{document}